# CRATERING OF SOIL BY IMPINGING JETS OF GAS, WITH APPLICATION TO LANDING ROCKETS ON PLANETARY SURFACES

P.T. Metzger, B.T. Vu, D.E. Taylor, M.J. Kromann, M. Fuchs, B. Yutko, and A. Dokos
*NASA, Kennedy Space Center, FL 32899*

C.D. Immer and J.E. Lane
*ASRC Aerospace, Kennedy Space Center, FL 32899*

M.B. Dunkel
*Security Assistance Corporation, Arlington, VA 22204*

C.M. Donahue
*Berry College, Mount Berry, GA 3014; University of Colorado, Boulder, CO 80309*

R.C. Latta III
*Embry-Riddle Aeronautical University, Daytona Beach, FL 32114*

## Abstract

Several physical mechanisms are involved in excavating granular materials beneath a vertical jet of gas. These occur, for example, beneath the exhaust plume of a rocket landing on the soil of the Moon or Mars. A series of experiments and simulations have been performed to provide a detailed view of the complex gas/soil interactions. Measurements have also been taken from the Apollo lunar landing videos and from photographs of the resulting terrain, and these help to demonstrate how the interactions extrapolate into the lunar environment. It is important to understand these processes at a fundamental level to support the on-going design of higher-fidelity numerical simulations and larger-scale experiments. These are needed to enable future lunar exploration wherein multiple hardware assets will be placed on the Moon within short distances of one another. The high-velocity spray of soil from landing spacecraft must be accurately predicted and controlled lest it erosively damage the surrounding hardware.

## Introduction

During the Apollo program, NASA investigated the blowing of lunar soil by rocket exhaust plumes in order to ensure safe landings for the Lunar Modules. The investigations were primarily theoretical (Roberts, 1963), experimental (Land and Scholl, 1969), or a combination of the two (Hutton, 1968), with very little numerical simulation relative to recent practice since most numerical capability has been developed during the intervening years. Another important source of information was the Surveyor program, in which a series of robotic spacecraft landed on the Moon, and especially the Surveyor III spacecraft when the Apollo 12 Lunar Module landed about 160 meters away from it. The Apollo astronauts returned portions of the Surveyor spacecraft to the Earth for analysis, and it was discovered that a thin layer of material had been scoured from the aluminum struts by the sandblasting of entrained soil in the Apollo Lunar Module's plume.  The surface of the Surveyor also had hundreds of pits (micro-craters up to 30 microns in diameter) from the impact of high-velocity soil particles (Cour-Palais, 1972), and a recessed location on the Surveyor's camera was contaminated with lunar fines (up to ~150 micron particles) where they had been blown through an inspection hole.  It was determined that the velocity of particles striking the Surveyor was between 300 m/s and 2 km/s (Brownlee, *et al.*, 1972).  At the time, the Surveyor spacecraft was already deactivated when the Lunar Module landed.  For functional hardware on the Moon, this sort of treatment will not be acceptable.  The scouring effects of the spray may ruin surface coatings, reflective blankets, and optics, and the injection of dust into mechanical joints may cause increased friction, jamming and mechanical wear.  In the future, when spacecraft return to the same location multiple times in order to build up a lunar base or to perform maintenance on scientific hardware, it will be necessary to prevent or block the spray.

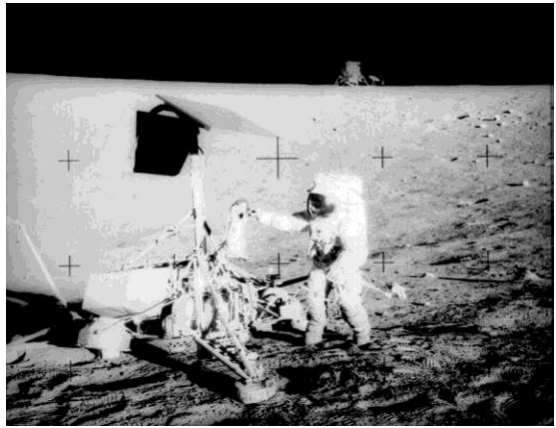

**Figure 1.  Surveyor III spacecraft with Apollo 12 Lunar Module on horizon**

Over the past decade, we have performed a number of investigations toward developing mitigation technologies for the spray of lunar soil and for the possibly deep cratering of martian soil.  A better understanding of the erosion and cratering physics were needed so that numerical simulations could be developed and effective mitigation strategies devised.  Toward that end we have performed experimental research of the dynamics of sand under impinging jets in conjuction with numerical simulations of the gas flow to improve the insight into the physics.  Within the allotted space, this brief paper will outline a few of the interesting results of this study.

## Mechanisms of Gas Moving Soil

During the Apollo era, researchers identified three physical mechanisms by which jets of gas can move soil.  These are:  Viscous Erosion (VE) described by Bagnold (1941), Bearing Capacity Failure (BCF) described by Alexander, *et al.* (1966), and Diffused Gas Eruption (DGE) described by Scott and Ko (1968).  VE is the sweeping away of the top layer of grains by the shear stress of a wall jet. BCF is the bulk shearing of the soil to form a cup beneath the stagnation pressure of a perpedicularly impinging jet, similar to cone penetration. DGE is an auxilliary effect, which occurs when the stagnation pressure drives gas into the pores of the soil, only to erupt carrying soil with it at another location or time. We have shown that there is actually a fourth mechanism, which we are calling Diffusion-Driven Flow (DDF). This occurs when the stagnation pressure drives gas through the soil so that the drag of the gas becomes a distributed body force in the soil, causing the soil to fail and shear. DDF and BCF both shear the soil in its bulk, but the former drives the soil tangentially to the crater's surface (radially from the tip and then up the sides), whereas BCF drives it perpendicularly to the crater's surface. In some cases, either DDF or BCF will dominate depending on whether the gas has sufficient time to diffuse through the soil. More generally, the gas diffuses only partially into the soil and the motion of the sand particles is intermediate to the two extremes (Metzger *et al.*, 2007). An example of this is shown in Figure 2, where a gas jet is forming a crater against a window in the sandbox where the motion of the sand can be observed.

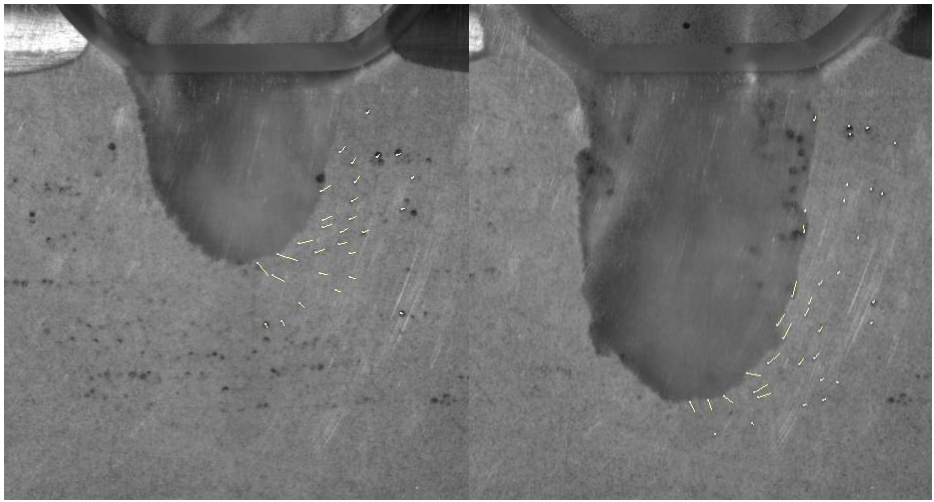

**Figure 2. Tracking individual sand grains beneath a growing crater**

Tests with supersonic plumes from solid rocket motors, such as in Figure 3, indicate that BCF is the primary mechanism when very large stagnation pressures are sharply focused onto a small region of the sand. In cases where the plume is highly underexpanded and produces a smaller pressure gradient on the soil's surface, it is expected that DDF could play a more significant role.

It should be noted that this sort of deep cratering does not occur in lunar landings because of the competence of the lunar soil and the absence of an atmosphere to collimate the plume and thus produce high pressure gradients. (However, it is believed that shallow scour holes did form in the final moments of some of the Apollo landings.) The benefit of these simple cratering experiments is that despite their dissimilarity to the lunar case they make it possible to identify the fundamental mechanisms of gas moving soil so that they can be coded into physics-based numerical simulations.

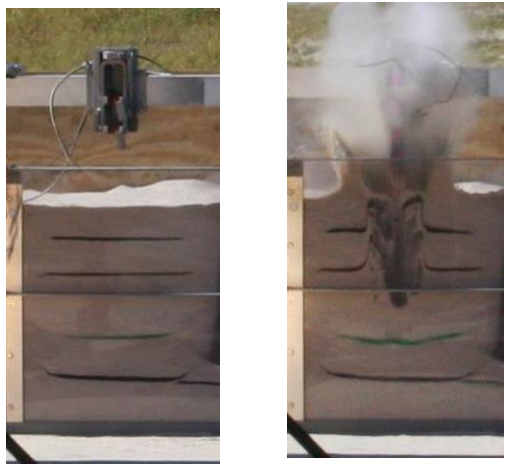

**Figure 3. Before and during the firing of a solid rocket motor in sand**

## Scaling of Cratering Dyanamics

In many cases of crater formation, the crater depth grows as the logarithm of time to good approximation for several decades as reported by Rajaratnam and Beltaos (1977) and others. This is shown in Figure 4 (Left). However, analysis of the crater volumes in these experiments (both the volume of removed sand and the volume of deposited "dunes" around the crater) as a function of time shows that the actual erosion rate is constant throughout this entire cratering process. The slowing rates of the crater depth and volume are due to the increasing recirculation of sand inside the crater as it widens and deepens, so that the same sand grain must be thrown into the air an increasing number of times before it finally escapes the crater.

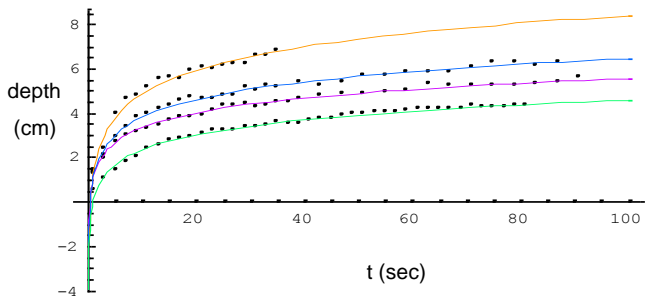
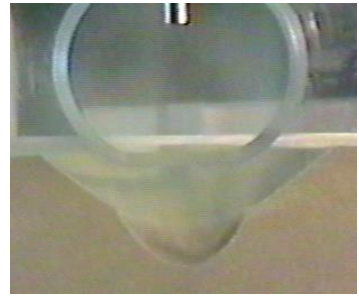

**Figure 4. Left: Crater depths versus time fitted to the logarithm of time. Right: Shape of crater, with inner crater at bottom of overall conical crater**

The constancy of the erosion rate is remarkable. It occurs because the jet forms and maintains a small "inner crater" of nearly constant dimensions at the bottom of the overall crater, even though the overall crater is growing in size. This is shown in Figure 4 (Right). All erosion occurs at the lip of the inner crater, where gas flow in contact with the inner surface of the crater is at its highest velocity. The circumference of this inner crater is determined by the drag of the gas holding the soil at a slope steeper than its angle of repose. If the inner crater erodes to have too large a circumference, then the drag per linear distance on its lip is diminished and sand from the slope of the outer crater collapse into the inner crater, shrinking its cicumference back to the original size. Thus, the inner crater automatically maintains a constant circumference, which ensures that the shear stress and erosion rate along its lip are also constant. Because of this, these experiments are a simple method to measure the scaling of that erosion rate on the inner lip.

Rajaratnam and Beltaos (1977) discovered that the asymptotic size of a crater scales with a unitless erosion parameter (called here the Rajaratnam number, *Ra*) equal to the densimetric Froude number multiplied by the jet diameter-to-height ratio,

$$Ra = \left[\frac{\rho_g V^2}{(\rho_s - \rho_g)gd}\right]^{1/2} \times \frac{D}{H}$$

where $\rho_g$ is the gas density, *V* is the jet velocity, $\rho_s$ is the sand mineral density, *g* is gravitational acceleration, *d* is the sand grain diameter, *D* is the jet diameter, and *H* is the height of the jet exit plane above the sand. Here, we have measured the rate of the dynamic processes of crater formation rather than the asymptotic size. For the physical parameters we have tested thus far we have found that the erosion rate does not scale with the Rajaratnam number. Instead, it is proportional to the dynamic pressure of the jet divided by its height ($\rho_g V^2 / H$). To date, we have not adequately tested the dependence on *d*, D, *g*, or $\rho_s$.

## Analysis of Apollo Landing Videos

Until the fully-integrated numerical simulations are completed, the best predictions for lunar soil erosion are still direct measurements taken from the Apollo landings. The morphology of the blowing dust can be measured by the shapes of the Lunar Module shadows that drape across it, as illustrated by the scale model in Figure 5 (Left), where a gas suspension beneath a plexiglass sheet serves as the "dust cloud". Analysis of these shadows in the landing videos indicates that the dust is blowing at an approximately 3 degree elevation angle away from the plume's stagnation region. Numerical simulation of the plumes indicates the dust velocity is on the order of 1 km/s, in agreement with Surveyor III data.

## Summary

The physics of the erosion and cratering processes have largely been determined by prior studies coupled with more recent work. Many simulations of lunar soil blowing (not presented here) have been performed, and new integrated flow codes with the cratering and erosion physics are being developed. An instrument capable of measuring these effects in lunar and martian landings is also being developed. Future work includes the benchmarking of the new flow codes by larger-scale experiments in vacuum chambers and by measurements during actual landings.

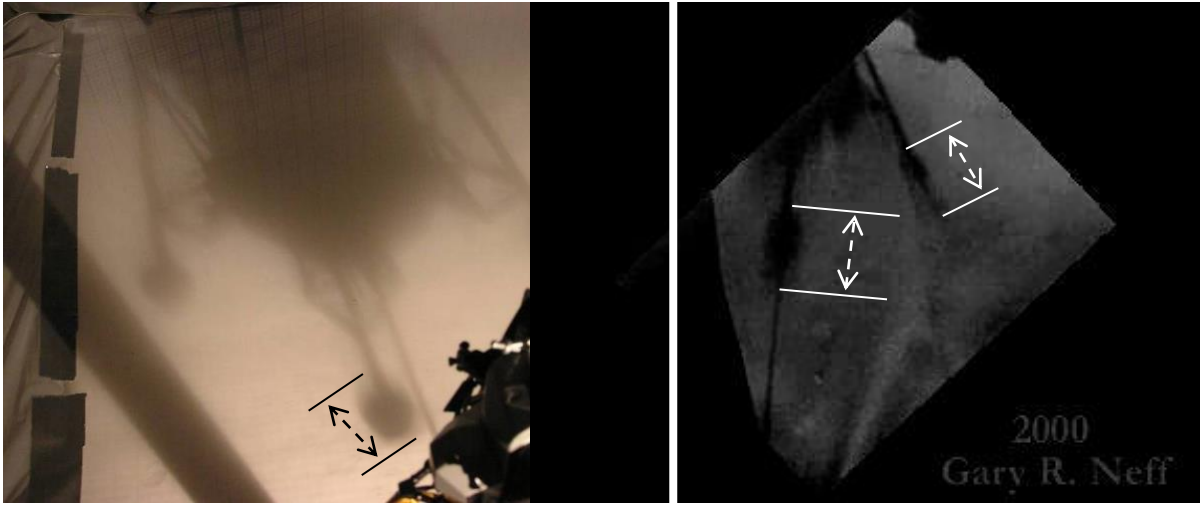

**Figure 5.  Lunar Module shadows on dust.  Left:  scale model  Right:  Apollo 17 (image credit:  Gary R. Neff)**

**References**

Alexander, J. D., W. M. Roberds, and R. F. Scott (1966), "Soil Erosion by Landing Rockets." Contract NAS9-4825 Final Report, Hayes International Corp., Birmingham, Alabama.

Bagnold, R.A. (1941), *The Physics of Blown Sand and Desert Dunes*, Dover, New York

Brownlee, D., W. Bucher, and P. Hodge (1972), "Part A.  Primary and secondary micrometeoroid impact rate on the lunar surface:  a direct measurement," SP-284, *Analysis of Surveyor 3 material and photographs returned by Apollo 12*, NASA, Washington D. C., 154–158.

Cour-Palais, B. G. (1972), "Part E. Results of Examination of the Returned Surveyor 3 Samples for Particulate Impacts," *Analysis of Surveyor 3 material and photographs returned by Apollo 12*, NASA, Washington D. C., pp 154–167.

Hutton, Robert E. (1968), "Comparison of Soil Erosion Theory with Scaled LM Jet Erosion Tests," NASA-CR-66704.

Metzger, Philip T., *et al.*, "Jet-induced cratering of a granular surface with application to lunar spaceports," *J. Aerospace Engineering* (to appear, 2007).

Rajaratnam, Nallamuthu, and Spyridon Beltaos (1977), "Erosion by Impinging Circular Turbulent Jets," *Journal of the Hydraulics Division*, **103**(10), 1191–1205.

Roberts, Leonard (1963), "The Action of a Hypersonic Jet on a Dust Layer," IAS Paper No. 63-50, *Institute of Aerospace Sciences 31st Annual Meeting,* New York, 1963.

Scott, Ronald F., and Hon-Yim Ko (1968), "Transient Rocket-Engine Gas Flow in Soil," *AIAA Journal* **6**(2), 258–264.